\documentclass{nature} 

\usepackage{setspace}
\usepackage{amsfonts}
\usepackage{amssymb}
\usepackage{amsmath}

\usepackage{float}
\usepackage{placeins}
\usepackage{graphicx}
\usepackage{bm}
\usepackage{braket}

\usepackage[export]{adjustbox}

\usepackage[nomarkers,nolists,figuresonly]{endfloat}

\usepackage[normal,normal,bf,labelformat=empty]{caption}

\usepackage[super,comma,sort&compress]{natbib}

\usepackage{hyperref}

\usepackage{verbatim}

\renewcommand{\i}{\mathrm{i}}
\renewcommand{\r}{\bm{r}}
\newcommand{\q}{\bm{q}}
\newcommand{\epar}{e^{\i \q_\parallel \cdot \r_\parallel}}
\newcommand{\ezp}{e^{\i q_j z}}
\newcommand{\ezm}{e^{-\i q_j z}}
\newcommand{\epzj}[1]{e^{\i q_{#1} z_j}}
\newcommand{\emzj}[1]{e^{-\i q_{#1} z_j}}
\newcommand{\wjp}[1]{w^+_{#1}}
\newcommand{\wjm}[1]{w^-_{#1}}
\newcommand{\vjp}[1]{v^+_{#1}}
\newcommand{\vjm}[1]{v^-_{#1}}
\newcommand{\qp}{\bm{q}_\parallel}

\usepackage{xcolor}

\title{In-situ spontaneous emission control of MoSe$_2$-WSe$_2$ interlayer excitons with near-unity quantum yield}
\author{Bo Han$^{1,\dag}$, Chirag Chandrakant Palekar$^{2,\dag}$, Sven Stephan$^{1,4}$, Frederik Lohof$^{3}$,Victor Nikolaevich Mitryakhin$^{1}$, Jens-Christian Drawer$^{1}$, Alexander Steinhoff$^{3}$, Lukas Lackner$^{1}$, Martin Silies$^{1,4}$, Bárbara Rosa$^{2}$, Martin Esmann$^{1}$, Falk Eilenberger$^{5,6,7}$, Christopher Gies$^{3,*}$, Stephan Reitzenstein$^{2,*}$ and Christian Schneider$^{1,*}$}
\begin{document}
\maketitle

\begin{affiliations}
 \item Institut für Physik, Fakultät V, Carl von Ossietzky Universität Oldenburg, Carl von Ossietzky Strasse 9-11, 26129 Oldenburg, Germany.
 \item Institut für Festkörperphysik, Technische Universität Berlin, Hardenbergstrasse 36, 10623 Berlin, Germany.
 \item Institute for Theoretical Physics and Bremen Center for Computational Material Science, Universität Bremen, Otto-Hahn-Allee 1, 28359 Bremen, Germany.
 \item Institute for Lasers and Optics, Hochschule Emden/Leer, Constantiaplatz 4, 26723 Emden, Germany.
 \item Institute of Applied Physics, Abbe Center of Photonics, Friedrich Schiller Universität Jena, 07745 Jena, Germany.
 \item Fraunhofer-Institute for Applied Optics and Precision Engineering, 07745 Jena, Germany.
\item Max Planck School of Photonics, 07745 Jena, Germany.
 \item[] $^{\dag}$ These authors contributed equally to this work
 \item[] $^{*}$ E-mails: \textcolor{blue}{gies@itp.uni-bremen.de; stephan.reitzenstein@physik.tu-berlin.de; christian.schneider@uni-oldenburg.de}
\end{affiliations}

\begin{abstract}
Optical resonators are a powerful platform to control the spontaneous emission dynamics of excitons in solid-state nanostructures. Here, we study a MoSe$_2$-WSe$_2$ van-der-Waals heterostructure that is integrated in a widely tunable open optical microcavity to gain insights into fundamental optical properties of the emergent interlayer charge-transfer excitons. First, we utilize an ultra-low quality factor open planar vertical cavity and investigate the modification of the excitonic lifetime as on- and off-resonant conditions are met with consecutive longitudinal modes. Time-resolved photoluminescence measurements reveal that the interlayer exciton lifetime can thus be periodically tuned with an amplitude of 110 ps. The resulting oscillations of the interlayer exciton lifetime allows us to extract a 0.5 ns free-space radiative lifetime and a quantum efficiency as high as 81 \%. We subsequently engineer the local density of optical states by introducing a spatially confined and fully spectrally tunable Tamm-plasmon resonance. The dramatic redistribution of the local optical modes in this setting allows us to encounter a profound inhibition of spontaneous emission of the interlayer excitons by a factor of 3.2. We expect that specifically engineering the inhibition of radiation from moiré excitons is a powerful tool to steer their thermalization, and eventually their condensation into coherent condensate phases. 
\end{abstract}


\section{Introduction}
Spontaneous emission is a fundamental process describing the radiative decay of the excited states. From a quantum electrodynamics viewpoint, it is triggered by vacuum fluctuations \cite{dirac1927quantum, welton1948some, milonni1984spontaneous, yokoyama1995spontaneous}. In order to tailor the spontaneous emission behavior, we need precise control of the electromagnetic density of states surrounding the emitters\cite{fermi1950nuclear}. Following the groundbreaking works of E. M. Purcell \cite{PhysRev.69.674.2}, controlling the spontaneous emission rate using a cavity that confines discrete modes has been widely demonstrated from atomic to low-dimensional solid-state systems \cite{goy1983observation, jhe1987suppression, kleppner1981inhibited,gerard1998enhanced, gerard2001inas}. In recent years, there has been significant research into two-dimensional semiconductors composed of transition metal dichalcogenides (TMDCs) due to their strong light-matter interactions and unique valleytronic properties \cite{wang2018colloquium}. Resulting from the quantum confinement effect and the reduced dielectric screening, the optical properties of atomically thin TMDC layers are governed by distinct exciton states with binding energy up to few hundred meV \cite{mak2010atomically, chernikov2014exciton, li2014measurement}. The outstanding optical properties of TMDCs render them ideal candidates for photonic applications and for investigations of solid-state cavity quantum electrodynamics \cite{schneider2018two}. 

The weak interlayer van der Waals interactions allow us to fabricate heterostructures with high versatility \cite{geim2013van}. When stacked TMDC heterobilayers (HBL) have a type-II band alignment, optical excitation of an individual monolayer results in layer-separated carriers and then generates interlayer excitons (iX) with permanent out-of-plane electric dipole moments ($\sim$0.5 nm$\cdot$e) \cite{seyler2019signatures, tran2019evidence, tang2021tuning,li2020dipolar}. Here, the iX has orders of magnitude longer lifetime (T$_1$ $\sim$ few ns) \cite{rivera2015observation} as compared to their intralayer counterparts featuring T$_1$ $\sim$ few ps\cite{fang2019control}. The radiative recombination of iX is furthermore influenced by the twist-angle dependent moiré periodicity across the 2D landscape, \textit{i.e.}, by the momentum space alignment of the K valleys of individual layers \cite{choi2021twist}, as sketched in Fig. \ref{figure 1}a. Besides varying the twist-angle, the ability to gain fully in-situ optical control of the excitonic dynamics is important for future photonic applications such as single-photon sources based on the quantum-dot alike moiré-trapped iX \cite{baek2020highly}. In this letter, we demonstrate the robust control of the iX lifetime in the weak coupling regime realized in a low-temperature open optical microcavity, which simultaneously provides valuable insights into their fundamental optical properties, including an ultra-high quantum efficiency exceeding 80\%.

\section{Results}

\subsection*{\textbf{Sample structure and experimental setups}}
\label{subsection_1}

\begin{figure}[t]
\centerline{\includegraphics[width=0.4\columnwidth]{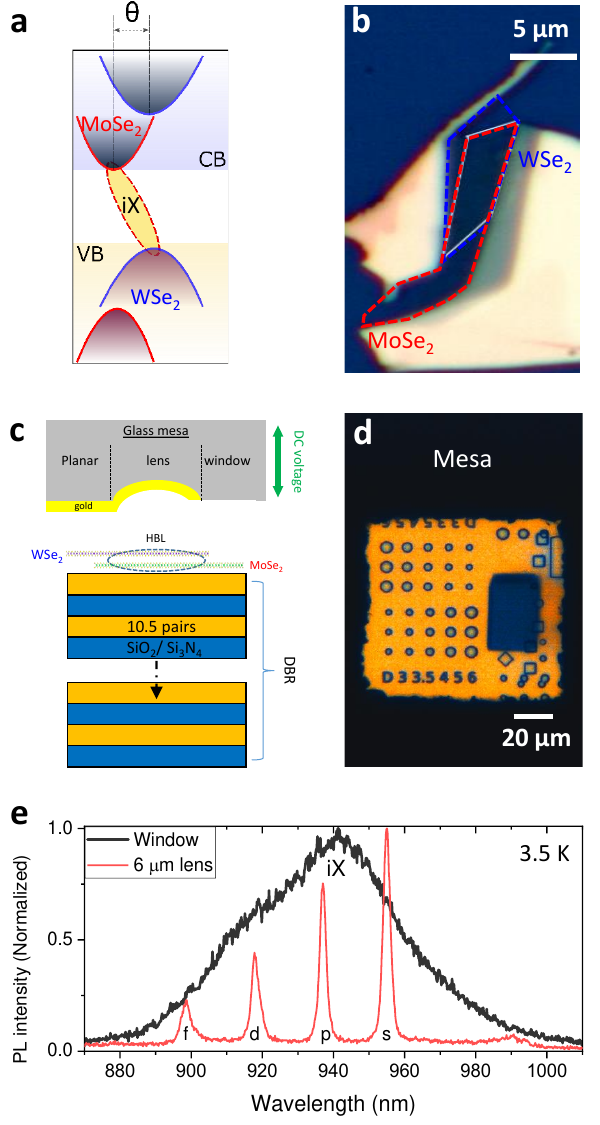}}
\caption{\textbf{Figure 1. Sample structures and open optical microcavity.} (a) Schematics of the type-II band alignment. The CBM and VBM are contributed by MoSe$_2$ and WSe$_2$, respectively. (b) Sample structure: WSe$_2$-MoSe$_2$-DBR (10.5 pairs of SiO$_2$/Si$_3$N$_4$). (c) Schematics of the open optical microcavity. The top mesa consists of a glass window area, gold-coated planar region and lenses. A maximum of 60 V DC voltage is used to drive the top mirror. (d) Top mirror: gold-coated mesa of 100x100 $\mu m^2$. The lens structures have same depth of 300 nm but different diameters of 3-6 $\mu$m. (e) PL spectra of iX at 3.5 K measured through the planar window and a 6 $\mu$m lens. The iX spectral distribution is as wide as 140 nm, as can be seen from the spectrum measured through the window (black). The spectrum measured through the lens (red) shows discretized \textit{s, p, d} and \textit{f} lens modes with Q-factors $\sim$400.}
\label{figure 1}
\end{figure}

Our sample is a $\theta$=1°$\pm$1°-twisted WSe$_2$/MoSe$_2$ HBL deposited on top of a dielectric Bragg reflector (DBR), see Fig. \ref{figure 1}b. As sketched in Fig. \ref{figure 1}c, the DBR contains 10.5 pairs of alternating SiO$_2$ (156 nm)/Si$_3$N$_4$ (114 nm), which results in a stop band centered at around 940 nm. The top mirror of the cavity is a 45 nm gold layer deposited by electron beam evaporation on top of a silica mesa of 100x100 $\mu$m$^2$ in size, whose microscope image is shown in Fig. \ref{figure 1}d. The surface of the mesa was pre-manufactured by focused ion beam of Ga$^+$ to produce several lens structures of hemispheric shape with uniform depth of 300 nm but different diameters. A microscope window of 20 x 40 $\mu$m$^2$ is also etched through the gold layer, enabling us to locate the TMDC heterostructure on the bottom DBR. Indeed, as we discuss later in this manuscript, even this window area forms an effective cavity of low quality factor (Q-factor). We can tune the cavity length by moving the top mirror vertically via a nano-positioner. A maximum of 60 V DC Voltage can be applied to the actuator to achieve a cavity detuning length up to approximately 0.6-1 $\mu$m, controlled with a sub-nm fine positioning resolution. The cavity is assembled in ambient condition and then loaded inside a dry closed-cycle cryostat with a base temperature of 3.5 K.

Photoluminescence (PL) measurements conducted in this paper are carried out by exciting the sample with 3 ps laser pulses at 750 nm from a mode-locked Ti:Sapphire laser. This excitation wavelength was chosen to generate iX efficiently in the heterobilayer, since it is nearly resonant with the MoSe$_2$ intralayer exciton at 3.5 K. Figure \ref{figure 1}e shows the comparison of the iX PL emissions collected through the window area and a gold-coated 6 $\mu$m lens. The PL emission centered at 940 nm from the window area manifests the explicit nature of the iX in our sample. The emission energy is lower than the excitonic resonances in individual monolayers, since it arises from the radiative recombination of the direct band gap in the HBL's K valleys that consists of valence band maximum (VBM) of WSe$_2$ and conduction band minimum (CBM) of MoSe$_2$. In addition, due to the combined effects of moiré reconstruction for slightly twisted monolayers with similar lattice constants and the inhomegeneity introduced during sample fabrication \cite{zhao2023excitons}, the PL profile is as wide as 140 nm spanning from 870 to 1010 nm. In contrast to this intrinsic iX spectral profile, the lens modes are more distinctive. The formation of localized Tamm-plasmon resonances in the DBR-vacuum-metal heterostructure yields Q-factors of each discretized transverse lens mode up to approximately 400.

\subsection*{\textbf{Tuning the iX lifetime in a low Q-factor cavity.}}

Our first experimental strategy towards in-situ control of the excitonic radiation dynamics is based on the effect, that even the transparent glass window in the top segment can form a resonant heterostructure. Although the cavity (Q-factor $\sim$24) is very lossy, the local electric field density on the DBR surface can be periodically tuned to influence the iX emission in the weak coupling regime (see Fig. \ref{figure 2}d). 

In similar configurations, the methodology to modify the emission dynamics by placing a dipole in the proximity of a dielectric interface was first discussed by Drexhage et al. \cite{drexhage1970influence}. It was later applied to study and quantify the spontaneous emission dynamics of quantum dots \cite{leistikow2009size, albert2010quantum, reitzenstein2011cavity}, excitons in TMDC monolayers \cite{fang2019control, zhou2020controlling, rogers2020coherent, horng2019engineering} and defects in hBN crystals \cite{nikolay2019direct}. However, thus far it has been sparsely implemented in a manner that allows for convenient in-situ tuning of the dielectric interface, which here enables us to explore exactly the same emitter ensemble throughout the entire study. Figure \ref{figure 2}a shows the stacked PL spectra of iX as a function of the cavity detuning. The spectra are collected through the planar glass window. The regions of the enhanced emission around 890 nm correspond to two longitudinal modes of low quality factor that tune through the broad PL emission of the iX. The mode numbers of 51st and 52nd longitudinal mode, and the effective cavity lengths are then determined by calculations using $L=q\dfrac{\lambda_{q}}{2}=(q+1)\dfrac{\lambda_{q+1}}{2}$, where $\lambda_{q (q+1)}$ are the wavelengths of the two adjacent longitudinal modes ($\textit{q, q+1}$) at the same DC Voltage (same cavity detuning). The effective cavity lengths are shown as the right axis of Fig. \ref{figure 2}a. 

We then performed time-resolved photoluminescence (TRPL) measurements on the iX emission with an avalanche photo diode (APD) at different cavity detuning from 0 to 60 V. Figure \ref{figure 2}b shows several TRPL spectra from 40 to 60 V, which display an acceleration of the iX decay process. It is important to note that the TRPL of iX cannot be fitted with a mono-exponential function, which would account for a single radiative decay channel only. We thus resort to a bi-exponential decay function\cite{choi2021twist}: $I(t)=Ae^{-t/\tau_1}+Be^{-t/\tau_2}$ to fit the TRPL spectra measured under different cavity detuning. We tentatively assign the faster decay ($\tau_2$) to the radiative recombination of the iX that is influenced by the cavity length and the slower decay process ($\tau_1$) to a non-radiative channel that is supposed to remain unaffected by the presence of a cavity resonance. And indeed the slower decay process is found to have an invariable value of $\tau_1$=12 ns after we performed the fitting procedure on all the recorded decay traces, while only $\tau_2$ changes with the cavity detuning. As shown in Fig. \ref{figure 2}c, $\tau_2$ oscillates periodically with a period of 40 V, matching with the period of the PL-intensity variation in the cavity detuning graph of Fig. \ref{figure 2}a. The PL maximums presented in Fig. \ref{figure 2}a correspond to the anti-nodes in the lifetime measurements. A sinusoidal fitting of the oscillation yields a $\tau_2$=2.2 ns, with an oscillatory amplitude of 110 ps. 

In the following, we calculate the free-space lifetime $\tau_0$ and quantum efficiency (QE) of iX. The long cavity length $\sim$23 $\mu$m can impose strong restrictions on the distribution of the electromagnetic modes inside the cavity. This is demonstrated by using the finite-difference in time-domain simulations (FDTD solutions, Lumerical). The schematics of the modelling box is shown in the left inset of Fig. \ref{figure 3}a, where the dipole is placed in the middle of a 5x5 $\mu$m$^2$ HBL on top of a DBR mirror and the detector is set 1.3 $\mu$m above the DBR surface for simulating the far-field emission pattern of the dipole. The refractive indices are taken for the wavelength of the observed mode. It is necessary to note that in stark contrast to the out-of-plane electric dipole moment of iX due to the layer-separated charge carriers, the nature of their optical dipole moments was actually determined as 99 \% in-plane \cite{sigl2022optical}, so that the in-plane radiative dipole in our simulation geometry is sufficiently justifiable. Figure \ref{figure 3}a and the right inset show the simulation results of the angle-dependent as well as the derived K-dependent intensity distribution of the cavity modes. The entire electromagnetic field intensity is predominated by a peak feature that can be fitted by a Gaussian function with a 4.5° full width at half maximum (FWHM). We account for this redistribution of the emission in the following modelling works and discard the weak intensities distributed at the high-angle flanks, which means that only the iX of center-of-mass wavevectors within the Gaussian profile are considered to substantially couple to the cavity.

The total decay rate for emitters embedded in the cavity can be expressed as $\Gamma_{tot}(r, \omega_0)=F_P\Gamma_0+\Gamma_\mathrm{nr}$, where $\Gamma_0$ and $\Gamma_\mathrm{nr}$ are the free-space and the non-radiative decay rate, respectively. The emission rate modification factor that is also the Purcell factor, $F_p=\frac{\Gamma_{cav}(r,\omega_0)}{\Gamma_0(\omega_0)}$, is defined as the ratio between the emission rate of iX in the cavity and the emission rate in free-space. The amplitude of this ratio changes alongside the cavity detuning and leads to the characteristic oscillating behavior as shown in Fig. \ref{figure 2}c. We calculate $F_p$ using the above-mentioned cutoff distribution. The details of the theoretical model can be found in the Method section and Supplementary information note 2 and 3.

We emphasize that the measurements do not show completely inhibited iX emission. This indicates a possible non-radiative decay channel that has finite decay rate similar to the total decay rate mediated by the cavity, and it can not be accurately disentangled from the bi-exponential fitting procedures that are previously applied to the TRPL traces. The total decay rate $\Gamma_{tot}$ is thus taken as $\Gamma_2$=1/$\tau_2$. In addition, the intercede of the total decay rate represents the hidden non-radiative decay rate where the emission is completely inhibited ($F_P=0$), and the slope is actually the free-space radiative decay rate $\Gamma_0$. Figure \ref{figure 3}b shows the linear fit of the $\Gamma_{tot}$ as a function of the calculated $F_P$, with the fitting parameters: $\Gamma_0=2.025\pm0.162$ ns$^{-1}$ and a second non-radiative decay rate $\Gamma_\mathrm{nr}^{(2)}=0.376\pm0.006$ ns$^{-1}$. We then acquire the free-space lifetime $\tau_0=1/\Gamma_0=0.5\pm0.04$ ns and the decay time of the second non-radiative channel $1/\Gamma_\mathrm{nr}^{(2)}=2.66\pm0.04 $ ns. We notice that the 2.66 ns decay time of the second non-radiative channel is indeed close to the cavity-mediated total decay time $\tau_2=2.2\pm0.11$ ns. After obtaining the $\Gamma_0$, we can calculate the quantum efficiency of iX by using the formula $QE=\frac{\Gamma_0}{\Gamma_0+\Gamma_\mathrm{nr}^{(1)}+\Gamma_\mathrm{nr}^{(2)}}$, where $\Gamma_\mathrm{nr}^{(1)}=1/\tau_1$ and $\Gamma_\mathrm{nr}^{(2)}$ are the decay rates of the two non-radiative channels that are obtained separately from the fitting of TRPL traces and our analytical approach in deducing the free-space lifetime. The quantum efficiency of iX is determined as 81.4$\pm$1.4 \% in case of near-resonant excitation of this 1°-twisted MoSe$_2$-WSe$_2$ HBL.    

To best fit the TRPL traces, an earlier work took into account only three radiative decay channels from the zero-momentum bright and gray iX, and also the phonon-assisted recombination of the spin-allowed finite-momentum iX which actually comprised a substantial spectral weight of the TRPL data \cite{forg2019cavity}. In contrast, the small twist angle of our sample facilitates the formation of near zero-momentum iX whose lifetimes are one order of magnitude shorter than those momentum-forbidden iX considered in the above-mentioned report. Therefore, the much faster non-radiative decay processes of 2.66 ns and 12 ns of the iX are feasible to be captured from our measurements. Given the combined effects of the small twist angle and the near-resonant excitation of the intralayer excitons in MoSe$_2$ monolayer, the layer-dependent charge transfer, energy relaxation and radiative recombination are more efficient in our sample, eventually resulting in the observed ultra-high Quantum efficiency.

\subsection*{\textbf{Inhibition of the spontaneous emission of iX coupled to the spatially localized Tamm plasmon modes}}

To further engineer the electromagnetic field distribution in our resonator, we resort to the integrated hemispheric and gold-coated lens structures in the top segment of the mesa (see Fig. \ref{figure 1}c). In conjunction with the planar bottom DBR which hosts the MoSe$_2$-WSe$_2$ heterostructure, this gives rise to zero-dimensional localized Tamm-plasmon resonances. In the following experiment, we reduced the effective cavity length to approximately 5 $\mu$m and utilized a hemispheric gold-coated lens with a diameter of 6 µm and a depth of 300 nm. Utilizing the open cavity implementation of the experiment, we scan the resonant Tamm-plasmon modes through the iX emission in the spectral range in 840-940 nm and observe clear variation of the PL intensities when optical resonances of the modes with Q-factors $\sim$ 400 are tuned through the iX emission profile, as shown in Fig. \ref{figure 4}a.  

To analyze the dynamics of our system, we performed TRPL measurements by collecting emission in the spectral range from 870 to 890 nm. Figure \ref{figure 4}b shows three representative TRPL spectra, collected for different cavity-iX detunings (or cavity energies E$_1$, E$_2$, and E$_3$ respectively). It is important to note, that the resonant cavity mode for energies E$_2$ and E$_3$ is within the spectral collection window, while in the case of E$_1$, emission of the cavity is not collected by the detector system. As the central result of our study, we observe a significant slow-down of the emission from our excitons in the case of a strongly detuned cavity. More specifically, from the TRPL decay traces in Fig. \ref{figure 4}b, we extract a decrease of $\tau_2$ from 2.3$\pm$0.03 ns in the resonant case to 7.3$\pm$0.3 ns in the off-resonant condition. Interestingly, the resonant decay time of $\tau_2$=2.3 ns at $E_3$ is in excellent agreement with the discussion in the previous section, and hints at only a weak on-resonant enhancement of spontaneous emission, while the significant slow-down is a clear-cut evidence of the inhibition of the radiative decay under off-resonant conditions. 

To provide a deeper understanding, we calculated the dynamics of a radiative dipole in our confined Tamm-plasmon structure using FDTD simulations. The cavity is modeled according to the experimentally determined values for the DBR as well as the concave top mirror. A linearly polarized dipole emitter is placed on top of the DBR and centered with respect to the top mirror. The dipole is surrounded by the heterobilayer which is approximated by two dielectric slabs. Symmetric and anti-symmetric boundary conditions are applied to reduce the computational effort. Thus, only the anti-symmetric modes with respect to the dipole axis can be excited. The wavelength dependent Purcell factor is retrieved from the power emitted by the dipole into the cavity. From the results of our calculation shown in the inset of Fig. \ref{figure 4}b, we can verify the observed modest acceleration of the spontaneous emission with a Purcell factor of only 1, while for the off-resonance spectral range it drops to a nearly constant value as low as 0.46, yielding the experimentally observed inhibition effect. 


\section{Discussion}

Our approach of controlling the dynamics of interlayer excitons in van-der-Waals heterostructures is of great importance, both, for future quantum optical studies and fundamental material related investigations. The periodic tuning of the spontaneous emission in widely tunable low-Q cavities allows to accurately extract the free-space exciton lifetime and quantum efficiency of complex samples, proving itself as an ultimately powerful characterization approach. The ultra-high experimentally observed quantum efficiency is of large importance for opto-electronic and quantum photonic applications of MoSe$_2$-WSe$_2$ interlayer excitons. In turn, engineering the inhibition of spontaneous emission in cavities can be directly exploited to enhance thermalization times of interlayer moiré excitons, which may become an important tool to promote their relaxation into coherent condensates.

\section{Methods}

\subsection{Sample fabrication} The HBL is fabricated by a dry-stamping method \cite{castellanos2014deterministic}. We determined the twist angle ($\theta$=1°$\pm$1°) by polarization-dependent second-harmonic generation measurements, see Supporting information note 1. The underlying DBR was grown on top of a silicon wafer by plasma-enhanced chemical vapor deposition. The glass mesa of the top mirror was cleaved by a micro-saw. The lens pits in the mesa surface and glass window are all etched by the focused ion beam of Gallium (FEI Helios 600i). The gold layer on the mesa was deposited in an electron beam evaporator (HHV Auto 500 lab coater).

\subsection{Optical microscopy} The open cavity was loaded into an attocube attoDRY1000 cryostat. All experiments were performed at a temperature of 3.5 K that is determined by the density of the surrounding helium bath. The piezo-based nano-positioners allow sub-nanometer tuning and stabilization of the cavity length. More detailed constructions of our cryogenic open-cavity can be found in this reference \cite{Jens2023monolayer}. The optical setup has a confocal geometry. The laser pulses (3 ps) for (TR)PL measurements were generated in a Coherent Mira Optima 900-F mode-locked Ti-sapphire laser that also synchronized with the time-tagger (quTAG) for the time-resolved measurements. Laser excitation and signal collection are achieved by a long work-distance lens objective (Thorlabs 354105-B, NA=0.6) that is placed on top of the cryogenic open cavity. The static PL signals were transmitted in a free-space beam path, collected by an Andor spectrometer (Shamrock SR-500i), dispersed by a 600 mm$^{-1}$ grating and finally recorded by a charge coupled device (CCD, iKon-M 934 Series). As for the TRPL measurements, we firstly utilized the spectrometer to disperse the static PL signal and truncated the spectral window of interest by using longpass and shortpass filters. The signal was then collected by a fiber-coupled zoom collimator. We found that this single mode fiber also functioned as a spatial filter and resulted in an improved signal-to-noise ratio, especially in case of the weak coupling between the Tamm-plasmon modes and iX. The output of the fiber is coupled with an APD (temporal resolution of 350 ps) whose electronic output was connected to the time-tagger. 

\subsection{Quantum efficiency extraction}
For extraction of $\Gamma_0$ and $\Gamma_\mathrm{nr}$, the ratio $\frac{\Gamma_{cav}(\omega_0)}{\Gamma_0(\omega_0)}$ is calculated as a function of the cavity length. We obtain the expression \begin{equation}
    \frac{\Gamma_{cav}(r,\omega_0)}{\Gamma_0(\omega_0)} = \frac{3n^3}{8 n_\mathrm{free}} \sum_{\sigma,\tau} \int_{0}^{\pi/2}\mathrm{d}\theta \, \sin(\theta) |\bm{u}^{\sigma,\tau}_{\theta}(z)\cdot \bm{e}_\parallel|^2 F_M(\theta),\label{eq:emission_enhancement}
\end{equation}
where $n$ = 1.2352 is the average refractive index between the two materials that surround the bilayer (air and SiO$_2$) and $n_{free}$ = 1 is the refractive index in the free-space case (or homogeneous medium). The mode functions $\bm{u}^{\sigma,\tau}_{\theta}(z)$ are calculated with a transfer matrix approach evaluated at the heterobilayer position. The mode functions are projected onto their in-plane component by $e_{||}$. The sum runs over two polarizations $\sigma$ and two propagation directions $\tau$ through the cavity structure, and we integrate over all emission angles. The distribution $F_M(\theta)$ is the angular emission distribution calculated for a dipole on the bottom DBR, \textit{i.e.}, the cutoff taken from Fig. 3a of the main text. Since the distribution is predominated by a Gaussian-shaped peak feature, it is then approximated by a Gaussian function $F_M(\theta)= \dfrac{1}{\sqrt{2\pi}\sigma} e^{\frac{-\theta^2}{2\sigma^2}}$
with FWHM = $2\sqrt{2ln(2)}\sigma = 4.5^\circ$.
A detailed derivation of Eq.~\eqref{eq:emission_enhancement} is given in the Supplementary information note 2 and 3.



\section {Data availability} Source data are provided with this paper.

\section{Reference}
\bibliography{nature-template}


\section*{Acknowledgments}
All authors gratefully acknowledge funding from the Deutsche Forschungsgemeinschaft (DFG) in the framework of SPP 2244 (funding numbers: Schn1376 14.1, Re2974/26-1 and Gi1121/4-1). B.H. acknowledges the Alexander von Humboldt-Stiftung for the postdoctoral fellowship grant and the support of the National Natural Science Foundation of China (Grant No. 12304012). M.S. and S.S. gratefully acknowledge funding from the Bundesministerium für Bildung und Forschung (BMBF) within the project tubLAN Q.0. M.E. acknowledges funding from the Carl von Ossietzky Universität Oldenburg through a Carl von Ossietzky Young Researchers' Fellowship. F.E. acknowledges support by DFG SFB 1375 (NOA) and BMBF FKZs 16K1SQ087K and 13XP5053A. The project was also partially funded by the QuanterERA II European Union’s Horizon 2020 research and innovation programme under the EQUAISE project, Grant Agreement No.101017733.

\section*{Author contributions} S.R. and C.S. conceived of the experiments. C.S. supervised the experiments. C.P and B.R. assembled the HBL sample. F.E. and M.E. prepared and manufactured the mesa. C.P. and B.H. performed the measurements with help from V.M. and L.L. B.H. and C.P. analyzed the experimental data. S.S. and M.S conducted the FDTD simulations. F.L., J-C.D, A.S. and C.G provided analytical model and theoretical interpretations to the experimental results. All authors discussed the results. B.H. F.L. and C.P. wrote the paper with inputs from all authors. 

\section*{Competing interests} The authors declare no competing interests.

\section*{Additional information}
\subsection{Supplementary information} is available for this paper.
\subsection{Correspondence} and requests for materials should be addressed to C.G., S.R. and C.S.

\begin{figure}[htbp!]
\centering\includegraphics[width=1\columnwidth]{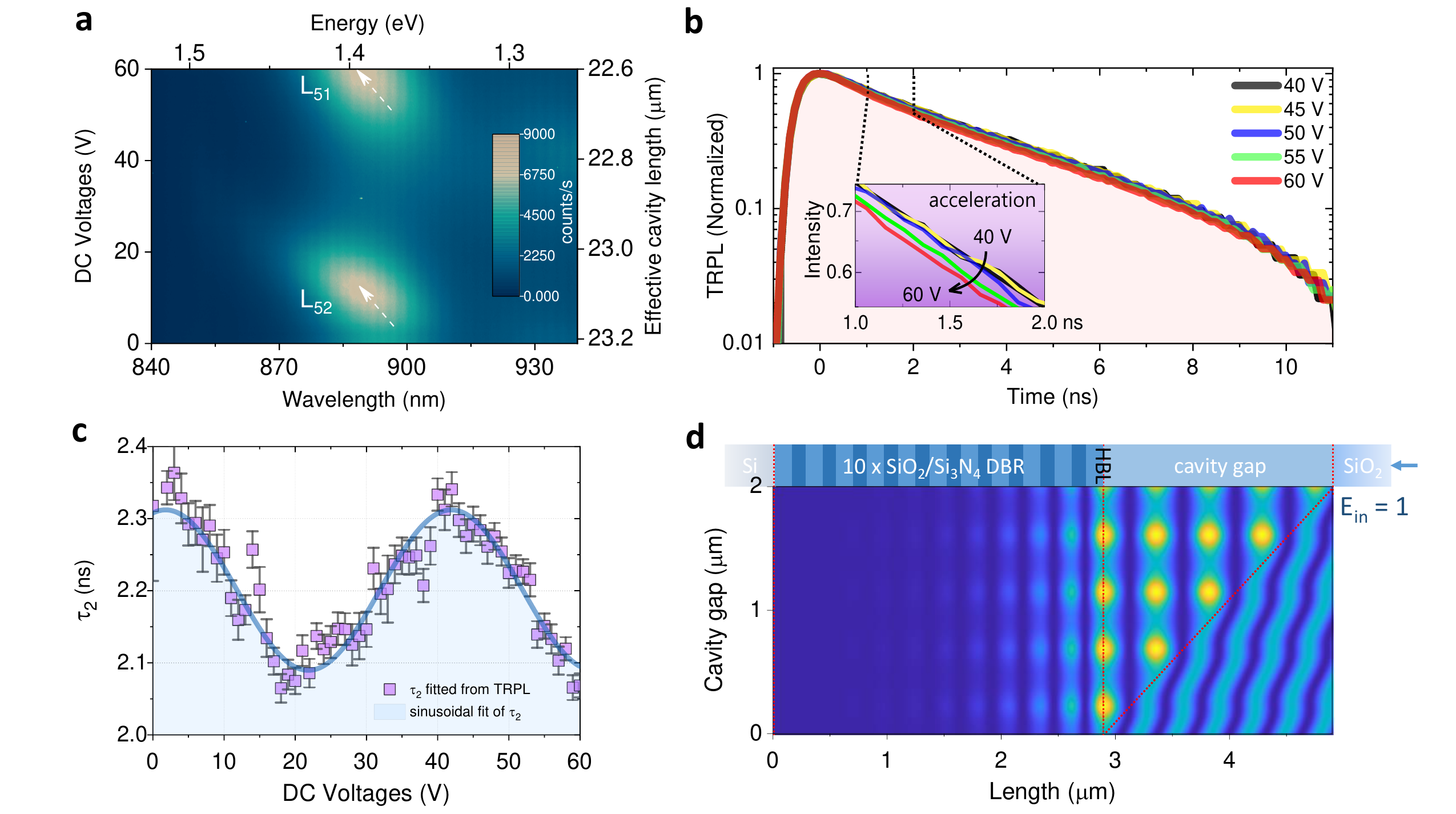}
\caption{\textbf{Figure 2. Tuning of the iX radiative decay in a planar glass-DBR cavity.} (a) Cavity detuning effects on the iX. The 51st and 52nd longitudinal modes enhance the PL intensity as they are tuned through the iX emission profile. (b) Selected time-resolved photoluminescence (TRPL) of the iX at different cavity detuning between 40-60 V. The acceleration of spontaneous emission can be seen from the inset of the enlarged region in 1-2 ns. (c) Tuning of the lifetime: the fitted $\tau_2$ of the faster decay process of iX oscillates with a period of 40 V, which corresponds to cavity length variation $\sim$400 nm. (d) Transfer matrix simulation of the absolute values of the electromagnetic field intensity in the glass-DBR cavity at perpendicular incidence as a function of the cavity gap, where the input field amplitude is fixed. The Q-factor is calculated as $\sim$24. The amplitude of the electromagnetic field density can change periodically with the cavity gap up to 23 $\mu$m (experimental values) to perturb the HBL, resulting in the oscillatory tuning of $\tau_2$, as shown in (c). }  
\label{figure 2}
\end{figure}

\begin{figure}[htbp!]
\centering\includegraphics[width=0.45\columnwidth]{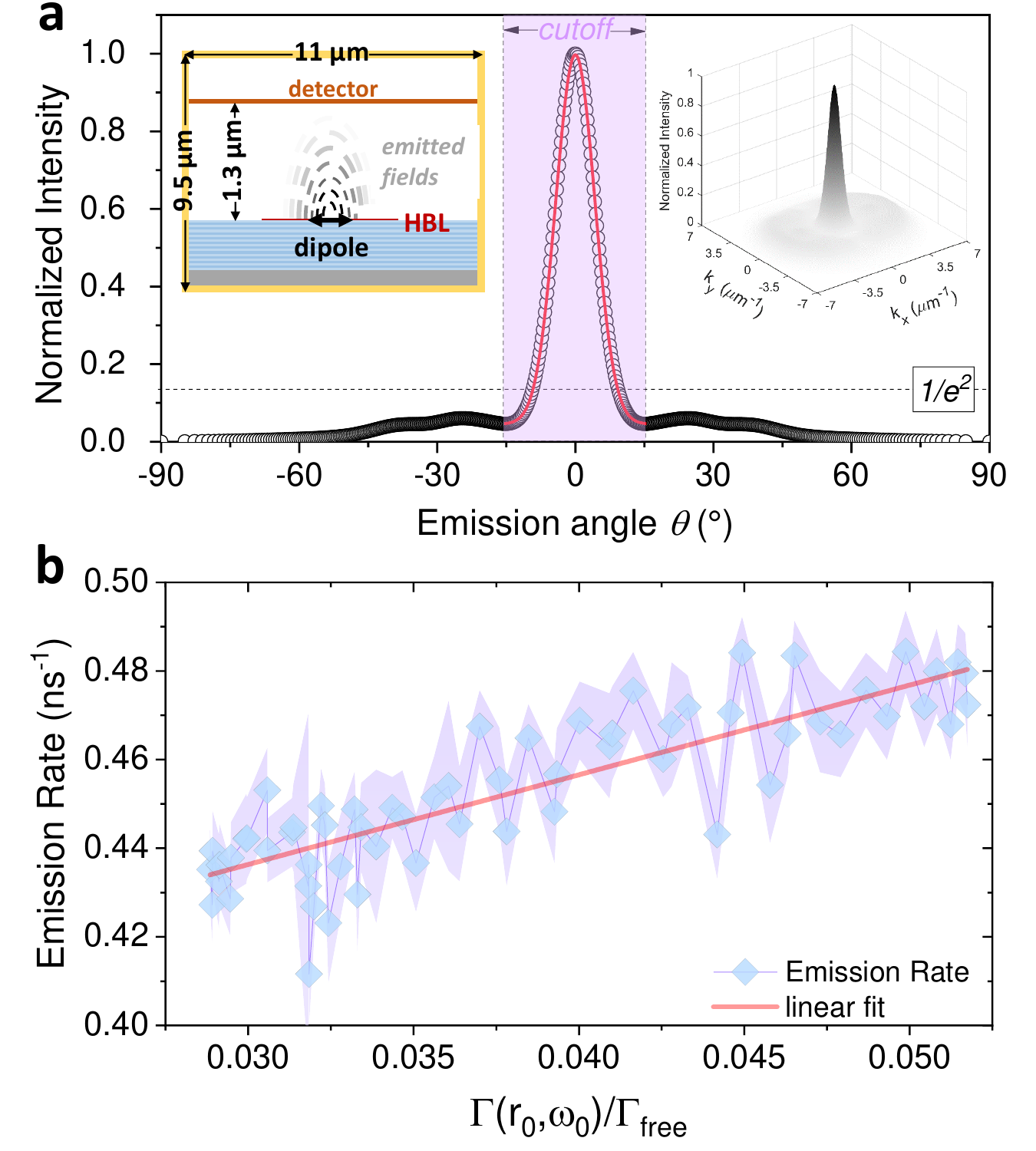}
\caption{\textbf{Figure 3. FDTD simulation and the fitting of iX decay rates.} (a) FDTD simulation results of the angle-dependent mode intensity distribution in the far-field. The magenta shaded region represents the cutoff which can be properly fitted by a Gaussian distribution function for further calculations of the emission rate modification factor $\Gamma(r_0, \omega_0)/\Gamma_0(\omega_0)$, $i.e.$, Purcell factor $F_P$. The red curve shows its fitting (FWHM$\sim$4.5°). For a better comparison, a dot line of $1/e^2$ of the maximum intensity is appended. It is substantially higher than the intensities distributed at the high-angle flanks, which justifies the selection of the Gaussian-shaped cutoff. Left inset: schematics of the FDTD simulation box. Right inset: the corresponding mode distribution in the quasi-particle momentum space. The k-vectors are converted from angles by using the formula $k_{\parallel}=2\pi\sin{\theta}/\lambda$, where $\lambda$ = 889 nm and $\theta$ is the emission angle relative to the normal direction, which is the same as the x-axis in the main graph. (b) Measured emission rate $\Gamma_\mathrm{tot}$ as a function of emission-rate modification $\Gamma_{cav}(r,\omega_0) / \Gamma_0(\omega_0)$ calculated from Eq.~\eqref{eq:emission_enhancement}. From a linear fit of the experimentally obtained decay rate we obtain radiative and non-radiative emission components: $\Gamma_0$=2.025$\pm$0.162 ns$^{-1}$ and $\Gamma_\mathrm{nr}^{(2)}$=0.376$\pm$0.006 ns$^{-1}$.} 
\label{figure 3}
\end{figure}

\begin{figure}[htbp!]
\centering\includegraphics[width=0.45\columnwidth]{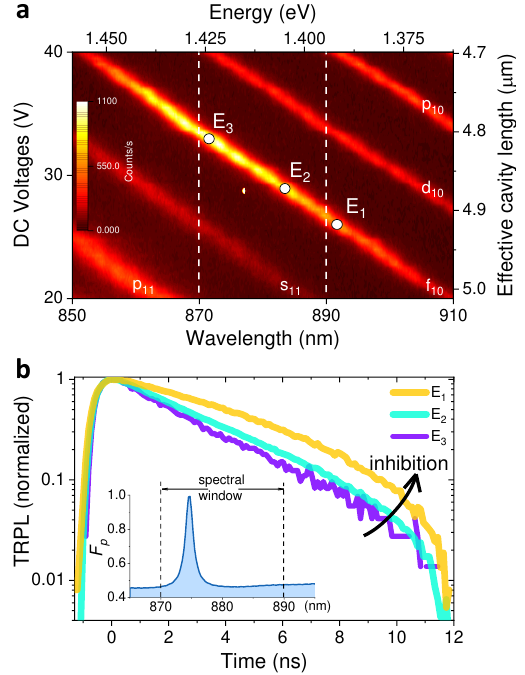}
\caption{\textbf{Figure 4. Inhibited spontaneous emission of iX coupled to a Tamm-plasmon-lens mode.} (a) Stacked PL spectra of the cavity scan using a 6 $\mu$m gold-coated lens. The discrete transverse lens modes are tuned through the iX emission profile. The 10th and 11th mode numbers are labelled in the graph. The vertical white dashed lines are the cutoff wavelengths of long-pass and short-pass filters used to confine a 870-890 nm spectral window for the following TRPL measurements. (b) TRPL measurements for emission energies at E$_1$, E$_2$ and E$_3$ that are labelled in (a). E$_1$ is outside the spectral window where iX dynamics are measured without the cavity effect. The TRPL at E$_2$ and E$_3$ show clearly faster decay rates as the lens mode resonance enters the spectral window. A bi-exponential fitting gives out a constant $\tau_1=18$ ns in these measurements, and a varying $\tau_2=7.3\pm0.3$ ns, $3.2\pm0.03$ ns and $2.3\pm0.03$ ns for emission energies at E$_1$, E$_2$ and E$_3$, respectively. The spectral window defined in (a) is also shown between the dashed lines. The inset shows the Purcell factor, F$_P$, calculated by the lumerical simulations, where the on- and off-resonance factors are 1 and 0.46, respectively.}
\label{figure 4}
\end{figure}

\newpage

\begin{center}

\Large\textbf{Supporting information for}

\large\textbf{In-situ spontaneous emission control of MoSe$_2$-WSe$_2$ interlayer excitons with near-unity quantum yield}

\normalsize\author{Bo Han$^{1,\dag}$, Chirag Chandrakant Palekar$^{2,\dag}$, Sven Stephan$^{1}$, Frederik Lohof$^{3}$,Victor Nikolaevich Mitryakhin$^{1}$, Jens-Christian Drawer$^{1}$, Alexander Steinhoff$^{3}$, Lukas Lackner$^{1}$, Martin Silies$^{1,4}$, Bárbara Rosa$^{2}$, Martin Esmann$^{1}$, Falk Eilenberger$^{5,6,7}$, Christopher Gies$^{3,*}$, Stephan Reitzenstein$^{2,*}$ and Christian Schneider$^{1,*}$}

\end{center}

\begin{affiliations}
 \item Institut für Physik, Fakultät V, Carl von Ossietzky Universität Oldenburg, Carl von Ossietzky Strasse 9-11, 26129 Oldenburg, Germany.
 \item Institut für Festkörperphysik, Technische Universität Berlin, Hardenbergstrasse 36, 10623 Berlin, Germany.
 \item Institute for Theoretical Physics and Bremen Center for Computational Material Science, Universität Bremen, Otto-Hahn-Allee 1, 28359 Bremen, Germany.
 \item Institute for Lasers and Optics, Hochschule Emden/Leer, Constantiaplatz 4, 26723 Emden, Germany.
 \item Institute of Applied Physics, Abbe Center of Photonics, Friedrich Schiller Universität Jena, 07745 Jena, Germany.
 \item Fraunhofer-Institute for Applied Optics and Precision Engineering IOF, 07745 Jena, Germany.
 \item Max Planck School of Photonics, 07745 Jena, Germany.
 
 \item[] $^{\dag}$ These authors contributed equally to this work
 \item[] $^{*}$ E-mails: \textcolor{blue}{gies@itp.uni-bremen.de; stephan.reitzenstein@physik.tu-berlin.de; christian.schneider@uni-oldenburg.de}
\end{affiliations}

\newpage

\section*{Supplementary note 1: Polarization dependent second-harmonic generation measurements}
\label{Supplementary note 1}

The twist angle between the monolayer of the heterostructure can be identified by polarization resolved second harmonic generation (SHG) measurements. Figure S1 shows SHG measurements for MoSe$_2$-WSe$_2$ heterostructure along with the extracted twist angle in heterostructure region. Both MLs were illuminated using linearly polarized laser at 1313 nm and SHG signal was measured at 656 nm. The intensity as function of excitation laser polarization is measured from the constituent MLs, which were illuminated with a linearly polarized light of a femtosecond mode-locked laser. In response, high intensity maxima are observed, as each maximum indicates the armchair direction on hexagonal crystal lattice of TMDC ML. Comparing the SHG response from the constituent MLs of the heterostructure, the twist angle can be determined as 1°$\pm$1°.

\section*{Supplementary note 2: Transfer matrix method for cavity mode calculation}
\label{Supplementary note 2}

Using the transfer matrix (TM) method, we derive electric field amplitudes inside a stack of planar sheets of non-magnetic, dielectric materials with dielectric constants $n_i$. The thickness of each slab is given by $\delta_i = z_i - z_{i-1}$, with the sheets oriented in the $x$-$y$-plane. The geometry is sketched in Fig. S2. We derive the field as a function of the angle of incidence (angle with normal direction, orthogonal to the planes).


\subsection{Basic equations}
The vector potential $\bm{A}(\r, t)$ satisfies the wave equation
\begin{equation}
    \left[\nabla^2 - \frac{n^2(\r)}{c^2}\frac{\partial^2}{\partial t^2} \right]\bm{A}(\r, t) = 0
\end{equation}
with stationary solutions $\bm{A}(\r, t) = A_0 \bm{U}_{\q \sigma}(\r)e^{\i \omega_q t}$ with the dispersion $\omega_q = c|\q|$ and $\q$ being the wave vector in vacuum and polarization index $\sigma$. $A_0$ carries the units of the vector potential while the mode functions are normalized according to
\begin{equation}
    \int\mathrm{d}^3\r n^2(\r) \bm{U}^\ast_{\q \sigma}(\r) \cdot \bm{U}_{\q' \sigma'}(\r) = \delta_{\q,\q'}\delta_{\sigma,\sigma'}\label{eq:mode_normalization}
\end{equation}
It is easy to see that the spatial component $\bm{U}_{\q \sigma}(\r)$ satisfies the Helmholtz equation
\begin{equation}
    \left[\nabla^2 + |\q|^2 n^2(\r) \right]\bm{U}_{\q \sigma}(\r) = 0.
\end{equation}

The system geometry sketched in Fig. S2 assumes an in-plane extension much larger than the wavelength and a dielectric function $n(\r) = n(z)$ piecewise constant along the $z$-direction. We thus make the ansatz $\bm{U}_{\q \sigma}(\r) = \bm{u}_{\q}^{\sigma}(z) \frac{1}{\sqrt{S}} \epar$, where $\q_\parallel$ is the in-plane component of the wave vector and $S$ is the quantization area. For the $z$-dependent part we get
\begin{equation}
    \left(\frac{\mathrm{d}^2}{\mathrm{d} z^2} + \left[|\q|^2 n^2(z) - |\q_\parallel|^2 \right] \right)\bm{u}_{\q}^{\sigma}(z) = 0.\label{eq:uqs}
\end{equation}
Since $n(z) = n_j$ in the layer with index $j$, we define $q_j = \sqrt{|\q|^2 n_j^2 - |\q_\parallel|^2}$ and find solutions for Eq.~\eqref{eq:uqs} of the form
\begin{equation}
    \bm{u}_{\q}^{\sigma}(z) = \bm{\varepsilon}_\sigma\left( A_j \ezp + B_j \ezm \right) ,\label{eq:z_mode_function}
\end{equation}
for each layer individually, where $\bm{\varepsilon}_\sigma$ indicates one of two polarization vectors ($\sigma=s$ or $p$).
It is important to note that in a planar geometry, we choose the coordinate system such that each mode travels in the plane of incidence, which is depicted in Fig.S3. We define it to be the $x$-$z$-plane. In each layer the light travels at an angle $\theta_j$ with respect to the normal direction. Thus the in-plane component and $z$-component are given by $|\q_\parallel| = |\q|n_j\sin(\theta_j)$ and $|\q_j| = |\q|n_j\cos(\theta_j)$, respectively. The in-plane component $|\q_\parallel| = q_x$ is constant throughout the structure due to Snell's law 
\begin{equation}
    n_1\sin(\theta_1) = n_2\sin(\theta_2),
\end{equation}
that is valid at a boundary of two media.
The amplitudes $A_j$ and $B_j$ are found by applying appropriate boundary conditions \cite{kira_quantum_1999}.

\subsection{Boundary conditions}
Boundary conditions are derived from Maxwell's equations and are given by continuity conditions for the tangential components of the electric field $\bm{E}$ and the magnetic flux density $\bm{B}$. Specifically, in the absence of surface charges and surface currents, the fields at the interface between two different materials (1 and 2) need to satisfy
\begin{equation}
    \bm{n} \times [\bm{E}_1-\bm{E}_2] = \bm{0}\label{eq:BD_E}
\end{equation}
\begin{equation}
    \bm{n} \times [\bm{H}_1-\bm{H}_2] = \bm{0}\label{eq:BD_H}
\end{equation}
\begin{equation}
    \bm{n} \cdot [\bm{D}_1-\bm{D}_2] = \bm{0}\label{eq:BD_D}
\end{equation}
\begin{equation}
    \bm{n} \cdot [\bm{B}_1-\bm{B}_2] = \bm{0}\label{eq:BD_B}
\end{equation}
where the electric field $\bm{H} = \bm{B}/\mu_0$ for non-magnetic materials, the displacement field $\bm{D} = \varepsilon_0\varepsilon_r\bm{E}$, and $\bm{n}$ is the normal vector to the interface. As $\bm{E}(\r) \propto \bm{U}_{\q \sigma}(\r)$ and $\bm{B}(\r) \propto \nabla \times \bm{U}_{\q \sigma}(\r)$, we apply the boundary conditions to the mode functions directly. We need to distinguish $s$ and $p$ polarization of the light.
\\
\paragraph*{\bf{s-polarization}}
For $s$-polarized light we use Eqs.~\eqref{eq:BD_E} and \eqref{eq:BD_H} for boundary conditions. The polarization vector (cf.~Fig.S3) points always along the $y$-direction $$\bm{\varepsilon}_s = \left(\begin{matrix}
            0 \\
            1 \\
            0
        \end{matrix} \right).$$
We thus find
\begin{equation}
    \sqrt{S} \,\bm{U}_{\q \sigma}(\r) = \bm{\varepsilon}_s\left( A_j \ezp + B_j \ezm \right)  \epar = \left(
    \begin{matrix}
            0                   \\
            A_j \ezp + B_j \ezm \\
            0
        \end{matrix} \right) \epar\label{eq:U_cont_s}
\end{equation}
and, noting that $|\q_\parallel| = q_x$ for our choice of coordinate system
\begin{equation}
    \sqrt{S} \,\nabla\times\bm{U}_{\q \sigma}(\r) =  \left(
    \begin{matrix}
            -\i q_j A_j \ezp + \i q_j B_j \ezm \\
            0                                  \\
            \i q_x A_j \ezp + \i q_x B_j \ezm
        \end{matrix}
    \right) \epar.\label{eq:cross_U_cont_s}
\end{equation}
The normal vector $\bm{n}$ at the boundary points always along the $z$-direction and therefore only the in-plane ($x$ and $y$) components of Eqs.~\eqref{eq:U_cont_s} and \eqref{eq:cross_U_cont_s} need to be continuous at the boundary.
\\
\paragraph*{\bf{p-polarization}}
For $p$-polarized light we use Eqs.~\eqref{eq:BD_E} and \eqref{eq:BD_D} for boundary conditions. The polarization vector (cf.~Fig.S3) depends on the angle of incidence $\theta$ in each layer and also on the propagation direction of the $z$-component ($\pm$) $$\bm{\varepsilon}^\pm_p = \left(\begin{matrix}
            \pm\cos(\theta) \\
            0               \\
            \sin(\theta)
        \end{matrix} \right).$$
We find, due to Eq.~\eqref{eq:BD_E}, that the in-plane component of
\begin{equation}
    \sqrt{S} \,\bm{U}_{\q \sigma}(\r) = \left(\bm{\varepsilon}^+_p A_j \ezp +\bm{\varepsilon}^-_p B_j \ezm \right)  \epar = \left(
    \begin{matrix}
            \cos(\theta_j) (A_j \ezp - B_j \ezm) \\
            0                                    \\
            \sin(\theta_j) (A_j \ezp + B_j \ezm)
        \end{matrix} \right) \epar\label{eq:U_cont_p}
\end{equation}
needs to be continuous. Furthermore with the refractive index satisfying $n^2 = \varepsilon_r\mu_r$ and using Eq.~\eqref{eq:BD_D}, the normal $z$-component of
\begin{equation}
    n^2(\r) \sqrt{S} \,\bm{U}_{\q \sigma}(\r)  = n^2(\r)\left(\bm{\varepsilon}^+_p A_j \ezp +\bm{\varepsilon}^-_p B_j \ezm \right)  \epar = \left(
    \begin{matrix}
            n^2_j\cos(\theta_j) (A_j \ezp - B_j \ezm) \\
            0                                         \\
            n^2_j\sin(\theta_j) (A_j \ezp + B_j \ezm)
        \end{matrix} \right) \epar\label{eq:U_cont_p2}
\end{equation}
needs to be continuous.

\subsection{Determining scattering coefficients}
The boundary conditions provide all the information needed to determine the complex coefficients $A_j$ and $B_j$. We note that each surface boundary results in two equations resulting in a set of $2(N+1)$ equations (cf.~Fig.S2). As we have $N+2$ different layers in our geometry and thus $2(N+2)$ unknown coefficients, we need to fix exactly two of the amplitudes to find a unique solution \footnote{We could also leave the two additional amplitudes open. This would result in an underdetermined system.}. From physical intuition we fix $A_0 = 1$ corresponding to the incident wave, and $B_{N+1} = 0$, as no light is scattered back from infinity. For another set of solution we exchange the roles of $A_0$ and $B_{N+1}$ corresponding to light incident from the opposite direction.
\\
\paragraph*{\bf{s-polarization}}
The in-plane ($x$,$y$) components in Eqs.~\eqref{eq:U_cont_s} and \eqref{eq:cross_U_cont_s} need to be continuous at the interface between two materials. The $z$-position between $n_j$ and $n_{j+1}$ is $z = z_j$. Inserting this and equating the expressions for both sides of the interface we obtain the following system of linear equations for the amplitudes
\begin{equation}
    0  = A_j \epzj{j} +  B_j \emzj{j} -  A_{j+1} \epzj{j+1} -  B_{j+1} \emzj{j+1},
\end{equation}
\begin{equation}
    0  = q_j A_j \epzj{j} -  q_j B_j \emzj{j} -  q_{j+1} A_{j+1} \epzj{j+1} +  q_{j+1} B_{j+1} \emzj{j+1},
\end{equation}
where $j\in \{0,\ldots,N\}$.
To determine the amplitudes, we solve the linear equation $M\bm{x}=\bm{b}$ with the solution vector $\bm{x} = (B_0, A_1, B_1, \ldots, B_N, A_{N+1})^\intercal$ and the right-hand side
\begin{equation}
    \bm{b} = (-A_0 e^{\i q_0 z_0}, -A_0 q_0 e^{\i q_0 z_0}, 0,0, \ldots, 0, 0, B_{N+1} e^{-\i q_{N+1} z_{N}}, -B_{N+1} q_{N+1} e^{-\i q_{N+1} z_{N}})^\intercal.
\end{equation}
For light traveling in the forward direction $A_0 = 1$ and $B_{N+1} = 0$. For light traveling in the backwards direction $A_0 = 0$ and $B_{N+1} = 1$. The determining matrix is, using $w^+_j = e^{\i q_j z_j}$, $w^-_j = e^{-\i q_j z_j}$, and $v^+_j = e^{\i q_{j+1} z_j}$, $v^-_j = e^{-\i q_{j+1} z_j}$:
\begin{equation}
    M = \left(\begin{matrix}
        \wjm{0}      & -\vjp{0}    & -\vjm{0}     & 0           & 0            & 0           & 0          & 0           & 0            & \cdots          \\
        -q_0 \wjm{0} & -q_1\vjp{0} & q_1\vjm{0}   & 0           & 0            & 0           & 0          & 0           & 0            & \cdots          \\
        0            & \wjp{1}     & \wjm{1}      & -\vjp{1}    & -\vjm{1}     & 0           & 0          & 0           & 0            & \cdots          \\
        0            & q_1 \wjp{1} & -q_1 \wjm{1} & -q_2\vjp{1} & q_2\vjm{1}   & 0           & 0          & 0           & 0            & \cdots          \\
        0            & 0           & 0            & \wjp{2}     & \wjm{2}      & -\vjp{2}    & -\vjm{2}   & 0           & 0            & \cdots          \\
        0            & 0           & 0            & q_2 \wjp{2} & -q_2 \wjm{2} & -q_3\vjp{2} & q_3\vjm{2} & 0           & 0            & \cdots          \\
        \vdots       & \vdots      & \vdots       & \vdots      & \vdots       & \vdots      & \vdots     & \vdots      & \vdots       & \vdots          \\
        0            & 0           & 0            & 0           & 0            & \cdots      & 0          & \wjp{N}     & \wjm{N}      & -\vjp{N}        \\
        0            & 0           & 0            & 0           & 0            & \cdots      & 0          & q_N \wjp{N} & -q_N \wjm{N} & -q_{N+1}\vjp{N} \\
    \end{matrix} \right).
\end{equation}
As the matrix $M$ is a banded matrix, the solution can be found more efficiently using a solver optimized for these kind of systems.
\\
\paragraph*{\bf{p-polarization}}
The \textit{in-plane} of Eqs.~\eqref{eq:U_cont_p} and the \textit{normal} component of Eqs.~\eqref{eq:U_cont_p2} need to be continuous at the interface between two materials. We obtain the equations
\begin{equation}
    0  = \cos(\theta_j)\left(A_j \epzj{j} -  B_j \emzj{j}\right)  -  \cos(\theta_{j+1})\left(A_{j+1} \epzj{j+1} -  B_{j+1} \emzj{j+1} \right),
\end{equation}
\begin{equation}
    0  = n_j \left(A_j \epzj{j} +  B_j \emzj{j}\right)  - n_{j+1}\left(A_{j+1} \epzj{j+1} +  B_{j+1} \emzj{j+1} \right),
\end{equation}
where in the second equation we used again Snell's law to eliminate the factor $n_j\sin(\theta_j) = n_{j+1}\sin(\theta_{j+1})$.
Analogous to the $s$-polarization we solve this linear system of equations for light traveling in the forward and backwards directions separately.

We note that in the case of light transmission \textit{through} a dielectric structure the approach can be simplified numerically by iteratively solving the system of equations layer by layer using $2\times 2$ matrices at each boundary from which the name \textit{transmission matrix} method derives. Our approach is more general and allows also to fix arbitrary coefficients \textit{inside} the structure, which is beneficial, e.g.~for calculation of waveguide modes that vanish outside the structure. 
\subsection{Mode normalization}
Given that many problems involve a summation over the infinite number of cavity modes, the question arises of proper mode normalizes in the numerical implementation. In analogy to calculations for free space modes one often wishes to perform a continuum limit of the form $\frac{1}{V} \sum_{\bm{q}}(\ldots)\rightarrow \frac{1}{(2\pi)^3}\int(\ldots)\,\mathrm{d}\bm{q}$, where $V$ is an abstract quantization volume for the modes. From the TM calculation we determine the mode functions $\bm{u}_{\q}^{\sigma}(z)$ as given in Eq.~\eqref{eq:z_mode_function}. The proper normalization is found by defining 
$\tilde{\bm{u}}_{\q \sigma}(z) = \sqrt{\frac{L_n}{A_{\q}}}\bm{u}_{\q}^{\sigma}(z)$ with 
\begin{equation}
    A_{\q} = \int_{-L_n/2}^{L_n/2}\mathrm{d}z\,n(z)^2|\bm{u}_{\q}^{\sigma}(z)|^2.\label{eq:normalization_integral}
\end{equation}
The length $L_n$ is the \textit{numerical} quantization volume that must be chosen much larger than the simulated cavity structure and large enough such that $A_{\q}/L_n$ converges to a constant value. With this normalization the mode function $\frac{1}{\sqrt{L}}\tilde{\bm{u}}_{\q \sigma}(z)$ for the cavity becomes completely analogous to the free-space mode function $\frac{1}{\sqrt{L}}e^{iq_z z}$ (in one dimension) to which the continuum limit can directly be applied. 

\subsection{Evaluation}
Here we provide a detailed description of input parameters to the TM calculation as well as details of the implementation. 
\begin{itemize}
    \item Free space angular frequency of the incident light $\omega_0 = 2\pi c_0/\lambda=$ with $\lambda = 890\,\mathrm{nm}$.
    \item The cavity-layer structure is depicted in Fig.\ref{figure 2}d. DBR structure consisting of 11 layers SiO$_2$ (156.42 nm) and 10 layers Si$_3$N$_4$ (114.37 nm). For the monolayers of MoSe$_2$ and WSe$_2$, we used a thickness of 0.7 nm, however these do not significantly influence the mode calculations. 
    \item Refractive indices of layer materials: $n$(Si) = 3.634, $n$(SiO$_2$) = 1.47, $n$(Si$_3$N$_4$) = 2.011, $n$(MoSe$_2$) = 5.51, $n$(WSe$_2$) = 4.23, $n$(Air) = 1
    
\end{itemize}
For the evaluation at different angles of incidence $\theta_0$ one needs to define the momentum-vector $z$-components $(q_0, q_1, \ldots, q_{N+1})$. Each value is determined via $q_j = |\q| n_j\cos(\theta_j) = \sqrt{|\q|^2 n_j^2 - |\q_\parallel|^2}$ with $|\q|=2\pi/\lambda$ and $|\q_\parallel| = |\q| n_0\sin(\theta_j)$ is the magnitude of the in-plane component of the wave vector, which is constant throughout the structure. For sufficiently steep angles, there will be total internal reflection inside certain layers which then act as wave guides. This can be interpreted in two different ways. First, if the angle of incident is larger than the critical angle, the value of $$\cos(\theta_{j+1}) = \cos\left(\arcsin(\frac{n_j}{n_{j+1}}\sin(\theta_j))\right) = \sqrt{1-\left(\frac{n_j}{n_{j+1}}\right)^2\sin^2(\theta_j)}$$ becomes imaginary. Another way to view this is to use the fact that $|\q_\parallel|$ stays constant due to Snell's law throughout the layers and as a result the square root becomes imaginary for $|\q| n_j < |\q_\parallel|$

Fig.~S4 shows the normalized mode functions $|\bm{u}^\sigma_{q_0\sin(\theta), q^\tau_0\cos(\theta)}(z)\cdot \bm{e}_\parallel|^2$ evaluated at the bilayer position and as a function of angle of incidence $\theta$ as well as the cavity length. The four panels represent individual calculations for different combinations of polarization (s and p) and transmission directions (forward, backwards). Horizontal lines indicate line cuts at a fixed cavity length, which are shown in Fig.~S5. For perpendicular angle of incidence ($\theta=0$) the two polarisations yield the same behavior as is expected from theory. 

\section*{Supplementary note 3: Emission rate modification}
\label{Supplementary note 3}

Following Ref.~\cite{stobbe_frequency_2009}, the Wigner Weisskopf theory for spontaneous emission starts with the state
\begin{equation}
    \ket{\psi} = c_e(t) \ket{c}\ket{0} + \sum_\mu c_\mu(t) \ket{v}\ket{1_\mu},
\end{equation}
where $\ket{c}$, $\ket{v}$ are the excited (c) and ground state (v) of an emitter and $\ket{0}$ and $\ket{1_\mu}$ are the photon states with zero photons (vacuum) and one photon in mode $\mu$, respectively.
Given this formulation the decay of the excited state $\ket{c}\ket{0}$ is given by the time evolution of the corresponding coefficient $c_e(t)$ 
\begin{equation}
    \dot{c}_e(t) = - \frac{q^2}{2m_0^2\varepsilon_0\hbar} |\braket{c|\bm{p}|v}|^2 \int_0^t \mathrm{d}t\,c_e(t') \int_{-\infty}^{\infty}\mathrm{d}\omega\,\frac{\rho(\bm{r}_0,\omega)}{\omega}e^{-\i (\omega-\omega_0)(t-t')}
\end{equation}
Assuming the expression $\frac{\rho(\bm{r}_0,\omega)}{\omega}$ varies slowly, it can be evaluated at $\omega_0$ and pulled out of the integral (cf.~Eq.~(8) in \cite{stobbe_frequency_2009}). In that case the remaining integral over $\omega$ gives a $2\pi\delta(t-t')$ and we arrive at $c_e(t) = c_e(0) \exp(-\Gamma_r t)$ with the radiative decay rate
\begin{equation}
    \Gamma_r = \frac{q^2\pi}{m_0^2\varepsilon_0\hbar} |\braket{c|\bm{p}|v}|^2 \frac{\rho(\bm{r}_0,\omega_0)}{\omega_0}\label{eq:gen_emission_rate}
\end{equation}
with the local projected density of states given by the sum of the mode functions normalized as in Eq.~\eqref{eq:mode_normalization} \cite{barnes_classical_2020}.
\begin{equation}
    \rho(\bm{r}_0,\omega_0) = \sum_\mu |\bm{U}_\mu(\bm{r})\cdot \bm{e}_\parallel|^2 \delta(\omega - \omega_\mu), \qquad \mu = (\qp, q_z, \sigma)
\end{equation}

\subsection{Emission rate in the cavity}
Here we calculate the emission rate modification at frequency $\omega_0$ from an iX distribution relative to emission from the same distribution into free space. If the iX of all momenta couple evenly to the photon modes the radiative emission rate is just proportional to the PLDOS as given in the previous section
\begin{equation}
    \Gamma_r(\omega_0, \bm{r}) \propto \sum_\mu |\bm{U}_\mu(\bm{r})\cdot \bm{e}_\parallel|^2 \delta(\omega_\mu - \omega_0), \qquad \mu = (\qp, q_z, \sigma).
\end{equation}
Here, $\omega_0 = \frac{c}{n}q_0$ and $\omega_\mu = \frac{c}{n}|\bm{q}| = \frac{c}{n} \sqrt{|\qp|^2 + q_z^2}$ and $n$ is the refractive index at the position of the iX. For the mode function we assume
\begin{equation}
    \bm{U}_\mu(\bm{r}) = \frac{1}{\sqrt{L}}\bm{u}_q^\sigma(z) \frac{1}{\sqrt{S}}\epar.
\end{equation}
Regarding proper normalization of the mode refer to Eq.~\eqref{eq:normalization_integral}. The modes $\bm{u}_q^\sigma(z)$ are determined from the TM calculation that assumes an infinite in-plane extension of the cavity structure. The finite extent of the cavity further modifies the coupling of the iX to the cavity modes focussing the emission to small emission angles. We account for this effect by an additional weighting factor $F_M(\qp)$ in the emission rate as discussed in the main text and method section.
Thus we find
\begin{equation}
    \Gamma_\mathrm{cav}(\omega_0, \bm{r}) \propto \sum_\mu |\bm{U}_\mu(\bm{r})\cdot \bm{e}_\parallel|^2 \delta(\omega_\mu - \omega_0) F_M(\qp).\label{eq:emission_gamma}
\end{equation}
Taking the continuum limit $\frac{1}{L}\sum_{q_z}\rightarrow\frac{1}{2\pi}\int\mathrm{d}q_z$ yields
\begin{align}
    \sum_\mu |\bm{U}_\mu(\bm{r}) & \cdot \bm{e}_\parallel|^2 \delta(\omega_\mu - \omega_0) F_M(\qp)\nonumber                                                                                                                                            \\
                                 & = \frac{1}{S}\sum_{\qp, \sigma} \frac{1}{2\pi}\int_{-\infty}^{\infty}\mathrm{d}q_z\,|\bm{u}_\mu(z)\cdot \bm{e}_\parallel|^2 \delta\left(\frac{c}{n} \sqrt{|\qp|^2 + q_z^2} - \frac{c}{n}q_0\right) F_M(\qp)\nonumber \\
                                 & = \frac{1}{S}\sum_{\qp, \sigma,\tau} \frac{n}{2\pi c} \, |\bm{u}^\sigma_{\qp,q_z^\tau}\cdot \bm{e}_\parallel|^2 \frac{q_0}{\sqrt{q_0^2-q_\parallel^2}}\theta\left(|q_0|-|\qp|\right) F_M(\qp)
    \label{eq:emission_at_omega0}
\end{align}
In the last step we used the well-known relation for the $\delta$-distribution
\begin{equation}
    \delta(g(x)) = \sum_{x_i} \frac{\delta(x-x_i)}{|g'(x_i)|}\qquad \mathrm{for}\qquad g(x_i)=0
\end{equation}
with $g(q_z) = \frac{c}{n} \left(\sqrt{\qp^2 + q_z^2}-q_0\right)$ and the roots of $g(q_z)$ given by $q_z^\tau = \pm\sqrt{q_0^2 + \qp^2}$ with $\tau \in \{+,-\}$ under the condition that $|q_0|\geq|\qp|$.
Next we take the continuum limit with respect to the in-plane momenta. This results in
\begin{align}
     & \sum_{\sigma,\tau} \frac{n}{(2\pi)^3 c} \int_{0}^{2\pi}\mathrm{d}\varphi\,\int_{0}^{q_0}\mathrm{d}q_\parallel \, |\bm{u}^\sigma_{\qp,q_z^\tau}(z)\cdot \bm{e}_\parallel|^2 \frac{q_0\,q_\parallel}{\sqrt{q_0^2-q_\parallel^2}}F_M(\qp)\nonumber \\
     & =\sum_{\sigma,\tau} \frac{nq_0^2}{4\pi^2 c} \int_{0}^{\pi/2}\mathrm{d}\theta \, \sin(\theta) |\bm{u}^\sigma_{\qp,q_z^\tau}(z)\cdot \bm{e}_\parallel|^2 F_M(q_0\sin(\theta))\nonumber                                                            \\
     & = \sum_{\sigma,\tau} \frac{n^3\omega_0^2}{4\pi^2 c^3} \int_{0}^{\pi/2}\mathrm{d}\theta \, \sin(\theta) |\bm{u}^\sigma_{q_0\sin(\theta), q^\tau_0\cos(\theta)}(z)\cdot \bm{e}_\parallel|^2 F_M(q_0\sin(\theta)),\label{eq:emission_general}
\end{align}
where we performed the integration over $\varphi$ and changed variables to the angle of incident $\theta$ via $q_\parallel = q_0 \sin(\theta)$ and $\mathrm{d}q_\parallel = q_0 \cos(\theta)\mathrm{d}\theta$.
\\

\subsection{Free space emission}
In the absence of the cavity the emitter emits into free space. In this case we evaluate Eq.~\eqref{eq:emission_general} for refractive index $n_\mathrm{free}$ and plane-wave mode functions, that satisfy the correct normalization condition
\begin{equation}
    \bm{u}^\sigma_{\qp,q_z^\tau}(z) = \frac{1}{\sqrt{n_\mathrm{free}^2}}e^{\pm \i q_z z}\bm{\varepsilon}_\sigma
\end{equation}
where $\bm{\varepsilon}_\sigma$ is the polarization vector. Also the weighting factor $F_M(\theta)$ is absent without the cavity. 
Inserting the plane-wave mode functions and summing over both, polarizations $\sigma$ and the two directions $\tau$, results in a free-space emission proportional to
\begin{equation}
    \Gamma_0(\omega_0) \propto \frac{n_\mathrm{free}\omega_0^2}{2\pi^2 c^3} \int_{0}^{\pi/2}\mathrm{d}\theta \, \sin(\theta) (1+\cos^2(\theta)) = \frac{2}{3}\frac{n_\mathrm{free}\omega_0^2}{\pi^2 c^3}.\label{eq:free_space_emission_general}
\end{equation}
We note that the sum over the two directions contribute a factor of 2 and the term $1+\cos^2(\theta)$ in the integral results from the summation over the polarizations, after projecting the $s$- and $p$-polarization vectors on the $x$-$y$-plane. 

\subsection{Final expression}
Finally the emission-rate enhancement from Eq.~\eqref{eq:emission_enhancement} mediated through the cavity is calculated via
\begin{equation}
    \frac{\Gamma_{cav}(r,\omega_0)}{\Gamma_0(\omega_0)} = \frac{3n^3}{8 n_\mathrm{free}} \sum_{\sigma,\tau} \int_{0}^{\pi/2}\mathrm{d}\theta \, \sin(\theta) |\bm{u}^{\sigma,\tau}_{\theta}(z)\cdot \bm{e}_\parallel|^2 F_M(\theta). \label{eq:emission_enhancement_total}
\end{equation}
We note that $\Gamma_{cav}(r,\omega_0)$ and $\Gamma_0(\omega_0)$ in Eqs.~\eqref{eq:emission_general} and \eqref{eq:free_space_emission_general} have the same proportionality constants in in front (c.f.~Eq.~\eqref{eq:gen_emission_rate}) which cancel in the final expression given above.

\begin{figure}[htbp!]
\centering\includegraphics[width=0.6\columnwidth]{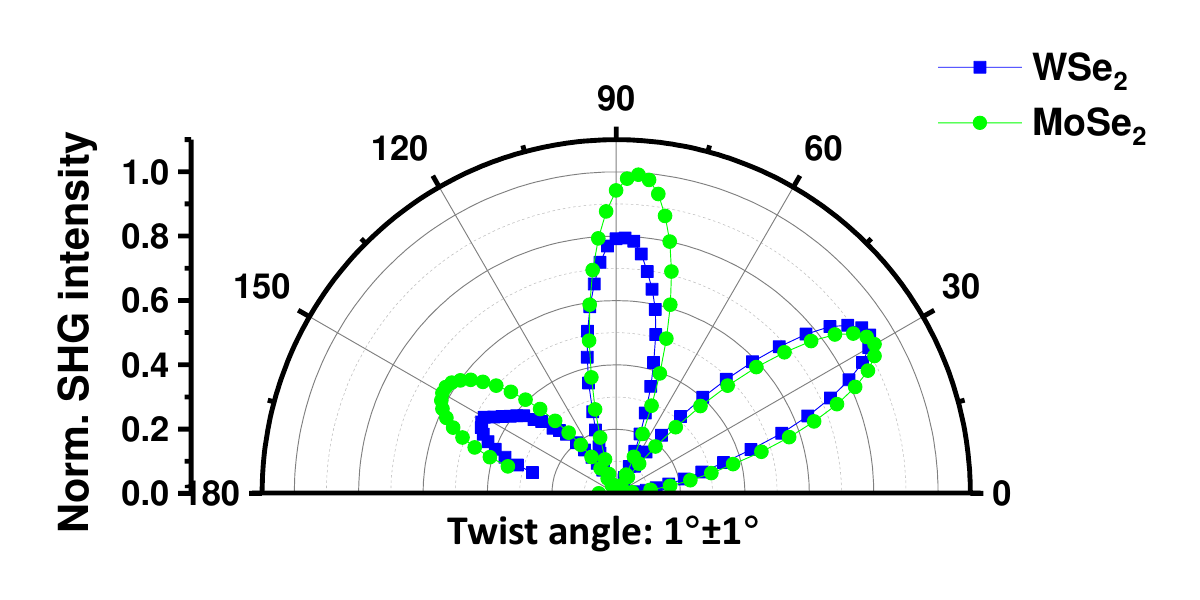}
\caption{\textbf{Figure S1. Second Harmonic Generation measurements.} }
\label{figure S1}
\end{figure}

\begin{figure}[htbp!]
    \centering
    \includegraphics[width=0.7\linewidth]{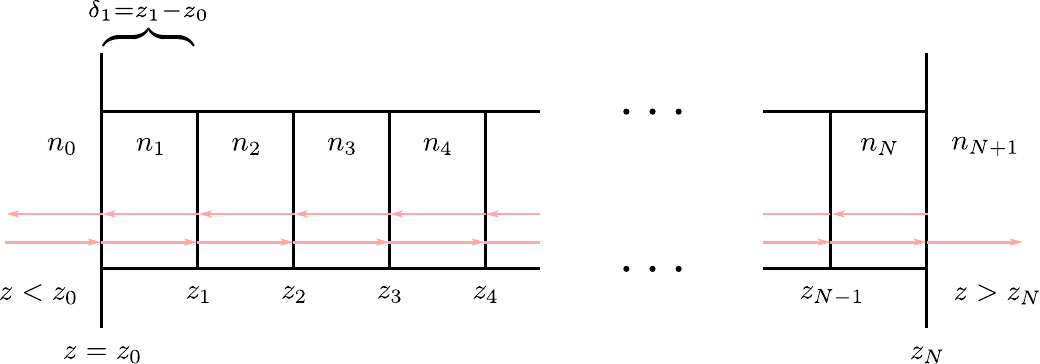}
    \caption{\textbf{Figure S2. Sketch of general system geometry in the transfer matrix approach.} The system is made up of layers of refractive index $n_i$ and thickness $\delta_i$. Outside the structure we assume homogeneous media with dielectric constants $n_0$ and $n_{N+1}$. Red arrows indicate propagation directions of the individual light-mode components in the case of forward emission through the structure. 
    }
    \label{fig:Geometry}
\end{figure}

\begin{figure}[htbp!]
    \centering
    \includegraphics[width=0.5\linewidth]{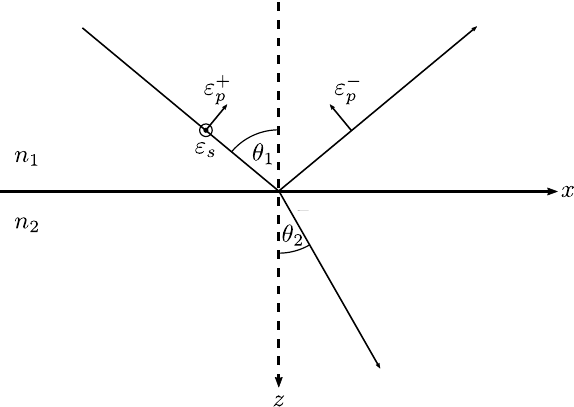}
    \caption{\textbf{Figure S3.}
        Refraction and reflection of light incident on a planar interface between two media at angle $\theta_1$. $\bm{\varepsilon}_s$ and $\bm{\varepsilon}^\pm_p$ indicate the polarization vectors for $s$ and $p$-polarized light.
    }
    \label{fig:plane_of_incidence}
\end{figure}

\begin{figure}[htbp!]
\centering\includegraphics[width=1\columnwidth]{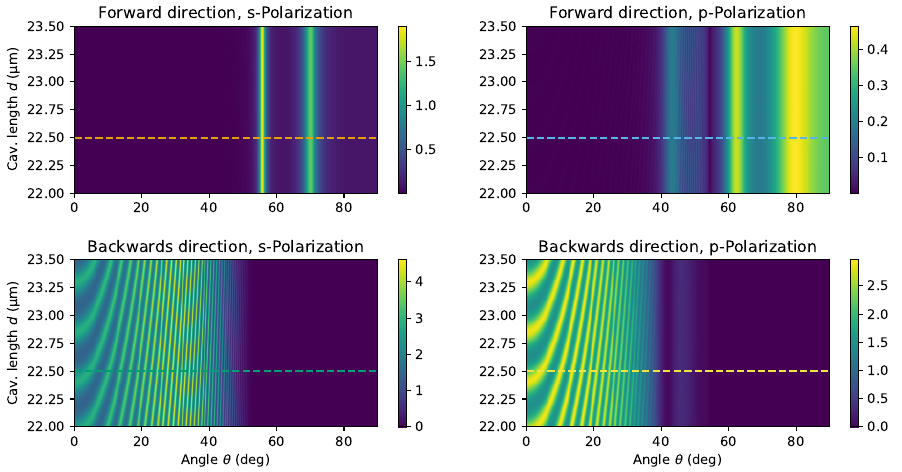}
\caption{\textbf{Figure S4.} Modes at emitter position as function of angle of incidence ($\theta$) and the cavity length for both directions and polarizations. Horizontal lines indicate the line cut shown in the Figure S3.}
\label{figure S4}
\end{figure}

\begin{figure}[htbp!]
\centering\includegraphics[width=0.8\columnwidth]{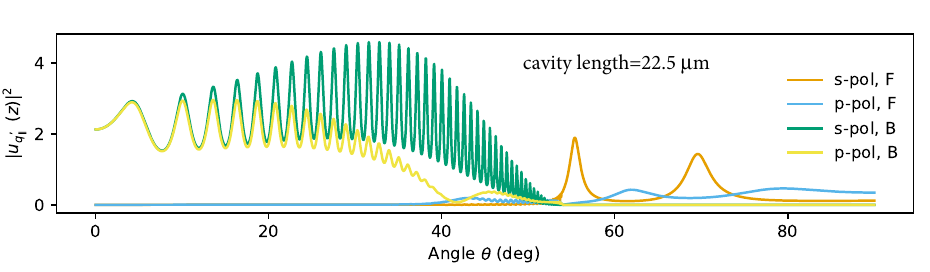}
\caption{\textbf{Figure S5} Modes at the fixed cavity length (d = 22.5 $\mu$m) as function of angle of incidence ($\theta$) for all combination of polarizations (s, p) and directions (F, B). Mode functions for both polarizations coincide for normal incidence ($\theta$ = 0).}
\label{figure S5}
\end{figure}

\end{document}